\documentclass[prl,twocolumn,longbibliography,nofootinbib]{revtex4-1}
\usepackage[utf8]{inputenc}
\usepackage[T1]{fontenc}
\usepackage{amsmath}
\usepackage{amsfonts}
\usepackage{amssymb}
\usepackage{graphicx}
\usepackage{xcolor}
\usepackage{enumitem}
\usepackage{physics}
\usepackage{overpic}
\usepackage[colorinlistoftodos, textwidth=\marginparwidth]{todonotes}
\usepackage[hidelinks]{hyperref}
\usepackage{grffile}
\usepackage{cleveref}
\newcommand{\fig}[1]{Fig.~\ref{#1}}
\newcommand{\Figs}[1]{Figures~\ref{#1}}
\newcommand{\Fig}[1]{Figure~\ref{#1}}
\newcommand{\Sex}{S_{\text{ex}}}
\newcommand{\Kex}{K_{\text{ex}}}

\begin{document}
    \author{Ian M. Douglass}\email{ian.douglass.42@gmail.com}
    \author{Jeppe C. Dyre}\email{dyre@ruc.dk}
    \author{Lorenzo Costigliola}\email{lorenzo.costigliola@gmail.com}
    \affiliation{{\it Glass and Time}, IMFUFA, Department of Science and Environment, Roskilde University, P. O. Box 260, DK-4000 Roskilde, Denmark} 
    \title{Complexity Scaling of Liquid Dynamics}
    \date{\today}
    \begin{abstract}
    According to excess-entropy scaling, dynamic properties of liquids like viscosity and diffusion coefficient are determined by the entropy. This link between dynamics and thermodynamics is increasingly studied and of interest also for industrial applications, but hampered by the challenge of calculating entropy efficiently. Utilizing the fact that entropy is basically the Kolmogorov complexity, which can be estimated from optimal compression algorithms [Avinery \textit{et al.}, Phys. Rev. Lett. \textbf{123}, 178102 (2019); Martiniani \textit{et al.}, Phys. Rev. X \textbf{9}, 011031 (2019)], we here demonstrate that the diffusion coefficients of four simple liquids follow a quasiuniversal exponential function of the optimal compression length of a single equilibrium configuration. We conclude that ``complexity scaling'' has the  potential to become a useful tool for estimating dynamic properties of any liquid from a single configuration.
    \end{abstract}
    \maketitle
    
Entropy defines the direction of time, a fundamental property that makes it an essential ingredient in the Gibbs and Helmholtz free energies determining whether or not a given process can take place \cite{rovelli,LLstat,reichl}. Half a century ago, Rosenfeld identified an entirely different property of entropy \cite{ros77}: Transport coefficients of simple liquids appear to be controlled by the excess entropy  per particle, $\Sex$, which is defined as the entropy in excess of an ideal gas at the same density and temperature ($\Sex < 0$ because any system is more ordered than an ideal gas). For a few simple liquids including the Lennard-Jones system, Rosenfeld demonstrated that $\tilde{D}\propto\exp(a\Sex)$ in which $a$ is a numerical constant and $\tilde{D}$ is the diffusion coefficient in macroscopically reduced units \cite{ros77,dyr18a}. Since then ``excess-entropy scaling'' has been shown to work also for many more complex systems, though often in a non-exponential form \cite{dyr18a}. This regularity is used, e.g., for estimating the viscosity and thermal conductivity of refrigerants and lubricating oils \cite{nov13,hop17,fou18}, the dynamics of electrolytes and silica melts \cite{bas11,goe11}, methane and hydrogen absorption in metal-organic frameworks \cite{jia15,liu16}, the viscosity of the Earth’s iron-nickel liquid core \cite{cao16}, separation of carbon isotopes in methane using nanoporous materials \cite{liu13,tia18}, etc. 

Excess-entropy scaling implies that the lines of constant excess entropy in the two-dimensional thermodynamic phase diagram are identical to the lines of constant reduced diffusion coefficient, reduced viscosity, etc. The isomorph theory predicts this for systems that obey hidden scale invariance \cite{IV,dyr14,sch14,dyr18a}, which however does not cover all known cases of excess-entropy scaling \cite{Bel19,bel19a}. 

It is textbook knowledge that entropy is an ensemble concept \cite{Fermi1937thermo,LLstat,reichl}. To determine a liquid's entropy $S$ in a simulation one typically, just as in experiments, employs thermodynamic integration starting from a state of well-known entropy, e.g., the dilute gas \cite{frenkel}. By involving several equilibrium simulations this method is tedious and computationally expensive, however, and the obvious question arises whether a more direct method exists for calculating a liquid’s entropy. For a simple liquid defined as a system of particles interacting via pair potentials \cite{han13}, a single snapshot allows for estimating the radial distribution function from which an important contribution to the excess entropy, the so-called two-body entropy $S_2$, can be determined \cite{net58,bar89}. There are also higher-order contributions to $S$ \cite{green52,bar89}, however, making the use of just $S_2$ an uncontrolled approximation. A method which in principle allows for determining the entropy from a single equilibrium configuration calculates first the chemical potential by Widom's random particle insertion method \cite{wid63}, but in practice this involves significant computation to avoid noisy data in the high-density (liquid) region of main focus here.

Excess-entropy scaling would be much more useful if entropy -- like pressure and temperature -- could be calculated easily from a single equilibrium configuration. Since this is not possible with current methods, one may ask whether some entropy proxy exists that can do the job well enough to be useful in practice. In other words: Does a quantity exist that is straightforward to evaluate for a single equilibrium configuration and which is, to a good approximation, in a one-to-one correspondence with the excess entropy? In 2019 two publications appeared answering this question to the affirmative, even for non-equilibrium states, by focusing on the \textit{Kolmogorov complexity} $K$ \cite{avi19,mar19}. Recall that for any set of data, $K$ is the length of the smallest algorithm that generates the data in question and subsequently terminates \cite{Cover_and_Thomas,Li_and_Vitanyi}. The connection to $S$ is that entropy is basically the logarithm of the number of states $N$ at a given energy (proportional to the density of states). An algorithm describing a typical one of these $N$ states to any given accuracy cannot be much shorter than $\ln(N)$ because generating different states requires different algorithms. On the other hand, specifying the Hamiltonian requires just a fixed length algorithm, and subsequently identifying the state in question to the given accuracy requires an added length of the algorithm varying as the logarithm of $N$. Consequently $K\sim\ln(N)\sim S$. 

Unfortunately, no algorithm exists that computes $K$ for an arbitrary string and then halts. This undecidability of the halting problem is a consequence of G{\"o}del's incompleteness theorem \cite{Cover_and_Thomas,Li_and_Vitanyi}. It thus might appear that Refs. \onlinecite{avi19,mar19} replaced the difficult task of estimating the entropy from a single configuration by the impossible one of calculating $K$. Physicists are pragmatic, however, and Refs. \onlinecite{avi19,mar19} suggested using compression algorithms to estimate $K$ by providing an upper bound, in this way estimating the entropy from a single configuration. 

We show below that compression algorithms are indeed good proxies for the excess entropy of simple liquids. This is done from simulations of several such systems, making it straightforward to predict the diffusion coefficient at a given thermodynamic state point from a single equilibrium configuration. First some details about the simulations and the compression of configuration files are given. Then the excess Kolmogorov complexity per particle, $\Kex$, is defined and a quasiuniversal correlation between $\Kex$ and the reduced diffusion coefficient $\tilde{D}$ is demonstrated. Finally, it is shown that there is a strong correlation between $\Sex$ and $\Kex$, thus validating complexity scaling via optimal data compression as useful whenever excess-entropy scaling applies. We here and henceforth use the term ``optimal data compression'' for the best method identified from the handful of tested compression algorithms.

\begin{figure}
        \centering
        \includegraphics[width=7cm]{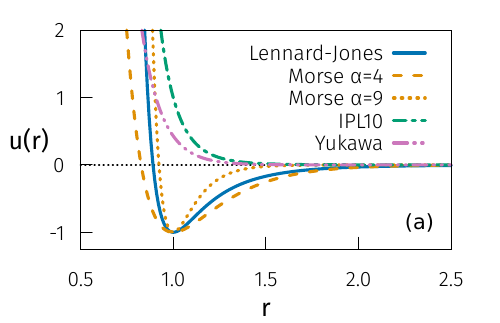} \\
        \includegraphics[width=7cm]{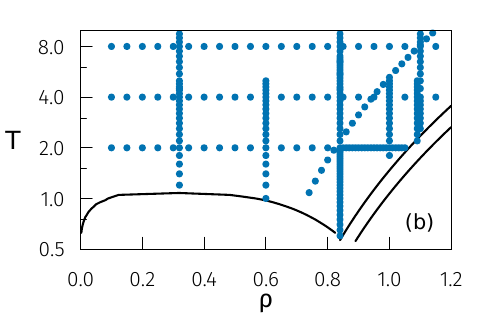}
        \caption{
        (a) Pair potentials studied. In this figure the Lennard-Jones potential is scaled for easy comparison with the Morse potential, while the IPL10 and Yukawa parameters are chosen to have a similar scale as they have no minimum to match.
        (b) Lennard-Jones phase diagram showing the state points simulated where T is the temperature and $\rho$ is the (number) density. The solid lines indicate phase boundaries.      }
        \label{fig:phasePots}
    \end{figure}

The model liquids considered are the standard Lennard-Jones (LJ) model \cite{lj24}, the inverse-power-law model with exponent $10$ (IPL10), the Morse model \cite{Morse1929}, and the Yukawa model \cite{yuk35}. The characteristic length and energy scales of the pair potentials are set to unity. The Lennard-Jones potential is thus given by $v(r) = 4 ( r^{-12} - r^{-6})$, IPL10 by $v(r)=r^{-10}$, the Morse potential by $v(r)=e^{-2\alpha(r-1)} - e^{-\alpha(r-1)}$ in which $\alpha$ between 4 and 9 were simulated, and the Yukawa potential by $v(r) = e^{-r}/r$ (Fig. \ref{fig:phasePots}(a)). All simulations were performed in the \textit{NVT} ensemble using a Nosé-Hoover thermostat and carried out using Roskilde University Molecular Dynamics (RUMD) \cite{RUMD} on RTX 2080 Ti GPUs. For the LJ, Yukawa, and IPL10 systems a wide region of the phase diagram was simulated, while for the Morse system isochores starting at the triple point were studied for several values of $\alpha$. As an example, Fig. \ref{fig:phasePots}(b) shows the state points simulated for the LJ systems. System sizes $N$, reduced time steps $\Delta \tilde{t}$, sample intervals, and cutoffs $r_{cut}$ are given in Table \ref{tab:simParams}.

\begin{table}
        \begin{tabular}{|c|cccc|}
            \hline
            System & $N$ & $\Delta \tilde{t}$ & Sample interval & $r_{cut}$ \\ \hline
            LJ & 32\,000 & 0.001--0.005 & $2^{15} \Delta \tilde{t}$ & $2.5$ \\
            IPL10 & 32\,000 & 0.001 & $2^{21} \Delta \tilde{t}$ & $2.5$ \\
            Morse & 32\,000 & 0.001--0.005 & $2^{15} \Delta \tilde{t}$ & $2.5$ \\
            Yukawa & 32\,000 & 0.001 & $2^{15} \Delta \tilde{t}$ & $4.3$\\ \hline
        \end{tabular}
        \caption{Simulation parameters for the four potentials considered. The columns report system size $N$, reduced time step $\Delta \tilde{t}$, sample interval, and cutoff radius $r_{cut}$. The time step at a state point is calculated as $\Delta t = \Delta \tilde{t} \sqrt{m / k_B T} / \rho^{1/3}$ from the reduced time step ($k_B=1$ and the particle mass $m$ is unity for all systems).}
        \label{tab:simParams}
    \end{table}

Before a compression algorithm can be applied to a configuration, it is crucial to find a data representation that is independent from the state point in question and from how the simulation was performed. Different compression algorithms will result in different final file sizes, and the Kolmorogov complexity $K$ is defined from the minimal-length algorithm. An upper bound for $K$ is determined by the compression algorithm resulting in the best compression. These individual steps have various implementations, with effects on the final compressed size as discussed later. The first step in the data preparation is to make the configurations easily comparable for the compression algorithms. This is done by scaling all configurations to unity density. At the same time we introduce a particle ordering using the locality-preserving 3D Hilbert curve \cite{Hilbert}. Particles are mapped to the nearest point on the curve such that all particles have a unique index; subsequently they are sorted so that consecutive particles are near to each other in 3D space.

The second step is to transform absolute positioning -- the $(x,y,z)$ positions within the simulation cell -- to relative positioning, i.e., the displacement from particle $i$ to $i+1$. In this way we remove any dependence from the simulation box used and improve the compression. We do this by using internal coordinates transforming the displacement from $(x,y,z)$ to $(r,\theta,\phi)$ where $r$ is the radial displacement from the previous particle, $\theta$ is the angle formed by the current and two previous particles, and $\phi$ is the four-particle torsion angle formed by the current particles and the three previous particles; this definition of $(r,\theta,\phi)$, which is standard in simulations of molecules, is illustrated in Fig. 2 of the Supplemental Material. Since the Hilbert-curve ordering is locality preserving, the distance between two particles with adjacent indices is almost always within the first or second coordination shell. This means that the distance is largely constrained to the inter-particle distance, which reduces the information needed to describe it. This transformation also exploits some of the ordering found in the simulated configuration, e.g., ``steric hindrance'' effects in $\theta$ (particles are unlikely to be folded back on themselves) and $\phi$. The transformations done to the initial coordinates in the two steps just described are lossless (reversible) and translate the simulation data into a series of displacement vectors. 

Following previous studies \cite{walraven2020entropy,zu2020information} we evaluate the difference between the optimal compression of a random configuration, i.e., an ideal gas configuration, and that of the configuration under study. Analogous to the definition of $\Sex$ we thus define the excess Kolmogorov complexity $\Kex$ as the difference between the Kolmogorov complexity of the configuration in question and that of a random configuration. The idea is that, what is physically relevant is how different the compressed configuration is from a random one of same density and temperature (and number of particles). This removes the issue of constant size components such as the compressor, any constants introduced, e.g., density scaling, and some dictionary length.

The most pivotal choice in the compression scheme is obviously that of the compression algorithm. The one used in this work was selected imposing the following five requirements: 
1) the results should be quantitatively reasonable, e.g., a lower entropy value should result in smaller sizes; 
2) the variance between different configurations of a given equilibrium simulation should be small enough to allow for accurate prediction from a single configuration; 
3) changing various parameters in the method (quantization size, etc.) should result in understandable changes;
4) the algorithm should have optimal compression, compatible with the criteria defining the Kolmogorov complexity itself; 
5) it should have a good time efficiency because choosing a method that takes several orders of magnitude longer obviates any potential benefit.

\begin{figure}
		\centering
		\includegraphics[width=7cm]{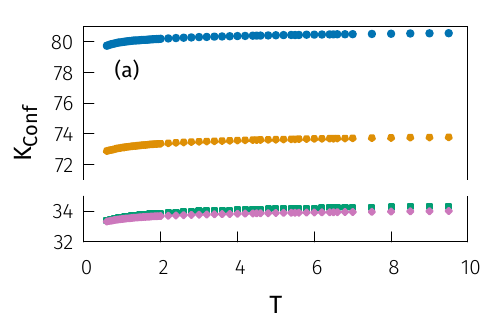} \\
		\includegraphics[width=7cm]{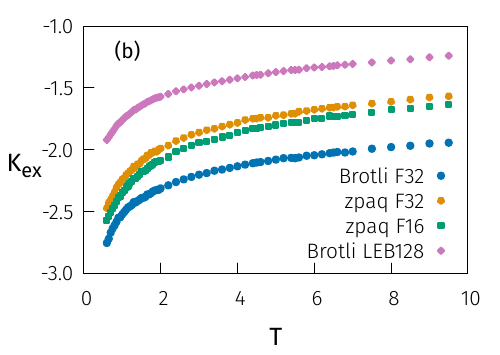}
		\caption{Comparison of compression methods along the Lennard-Jones $\rho = 0.84$ isochore as a function of temperature. In (a) the absolute bits/particle size, denoted as $K_{\rm Conf}$, are shown for four compression methods, while (b) shows data for the same methods plotting the difference between a simulated configuration and a random one.}
		\label{fig:deltaComparisons}
	\end{figure}

Figure \ref{fig:deltaComparisons} compares results for several compression methods. For simplicity of presentation only data regarding the LJ system are shown here, but similar results apply for the other systems studied. The four compression methods compared in the figure are two version of Brotli (LEB128 and F32) \cite{alakuijala2018brotli,brotliGithub} and two versions of zpaq (F16 and F32) \cite{mahoney2009incremental,zpaqGithub}. These versions of Brotli and zpaq utilize different representation for the data (integer quantization or not, and different choices of the float precision); details about the compression algorithms studied are given in the Supplemental Material. In (a) we show per-particle size in bits of the compressed data along the $\rho = 0.84$ isochore of the LJ system, while (b) shows the difference to a random/ideal-gas configuration. 

\Fig{fig:deltaComparisons}(b) shows that the general shape of $\Kex$ remains the same even after changes in the numerical representation and the compression method used, showing a robustness to the approach taken. The compression results have a monotonous dependence upon temperature at fixed density, as expected for the thermodynamic excess entropy. For the results presented below, zpaqF32 is used because it obeys all above five requirements. Importantly, while the standard deviation of a single sampling of a configuration is approximately 2\% of the mean, we can re-sample a single configuration using different starting points of the Hilbert curve to recover the same distribution as sampling once from multiple configurations (Fig. 1 of the Supplemental Material). This means that only a single configuration is needed to reliably estimate $\Kex$. 

The use of a 32-bit floating-point value provides an excess of resolution---at least as accurate as the simulation (run at 32-bit precision on GPUs) itself---to dispel any doubts as to under-representing the data. This compression provides a better raw compression over Brotli (in concordance with the upper-bound concept of Kolmogorov complexity), as well as better handling of the 16-bit floating point representation. The downside to this choice is that this compression method is more of a black box compared to others. 

\begin{figure}
	\centering
	\includegraphics[width=7cm]{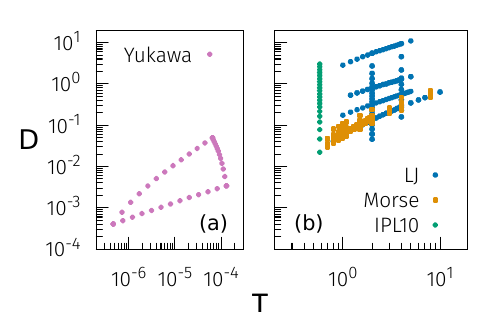} \\
    \includegraphics[width=7cm]{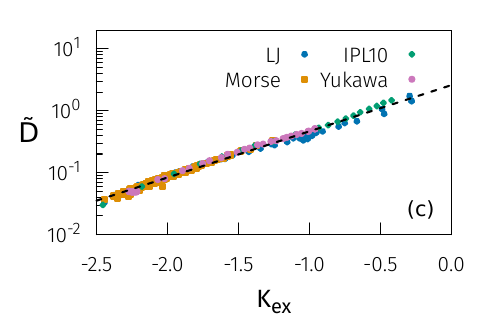}
	\caption{Diffusion coefficients $D$. 
        (a) shows $D$ for the Yukawa system as a function of the temperature along three lines in the thermodynamic phase diagram (one of constant density, one of constant inverse average interparticle distance over temperature, and one isomorph \cite{vel15}).
        (b) shows similar data for the LJ, IPL10, and Morse potentials, along isotherms and isochores.
        (c) gathers all data in a plot of the reduced diffusion coefficient $\tilde{D}$ versus $\Kex$ estimated from the zpaqF32 compression algorithm, demonstrating a quasiuniversal exponential functional dependence (black dashed line: $\tilde D\propto \exp(c K_{ex})$).}	\label{fig:KexSex}
\end{figure}

To test the predictive power of $\Kex$ we determined the diffusion coefficient $D$ at several state points for each of the four pair potentials. \Figs{fig:KexSex}(a) and (b) show the diffusion coefficients versus temperature for all state points simulated, while (c) plots the reduced diffusion coefficient defined by $\tilde{D}\equiv\rho^{1/3}\sqrt{m/k_BT}D$ \cite{ros77,IV} ($m$ is the particle mass set to unity in the simulations and $k_B$ is the Boltzmann constant) versus $\Kex$ estimated from zpaqF32. We find an exponential quasiuniversal functional dependence. This is our main result, demonstrating that the diffusion coefficient can be reliably estimated from the compression of a single equilibrium configuration.

\begin{figure}
    \centering
    \includegraphics[width=7cm]{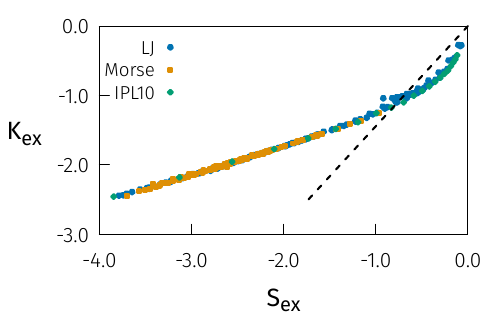}
    \caption{Correlation between $\Kex$ estimated from the zpaqF32 compression algorithm and the excess entropy $\Sex$. The dashed line represents idealised scaling.}
    \label{fig4}
\end{figure}

Assuming that optimal data compression provides a good estimate of $\Kex$ and thereby of the excess entropy $\Sex$, the finding of \fig{fig:KexSex}(c) is consistent with excess-entropy scaling as well as the quasiuniversality of simple liquids (the fact that structure and dynamics are very similar in reduced units) \cite{ros99,dyr16}. A strong correlation between $\Kex$ and $\Sex$ is confirmed in \fig{fig4} in which $\Kex$ is obtained as above and $\Sex$ is obtained from the Thol \textit{et al.} equation of state \cite{thol2015equation,thol2016equation} for the LJ system and by thermodynamic integration in the other cases ($\Sex$ data are not given for the Yukawa system because the 
simulation points in this system do not lie on standard integration paths). The correlation between $\Kex$ and $\Sex$ is very good; in fact it is linear over most of the range investigated in this work, meaning that $\Kex = \alpha \Sex + \beta$. Here, $\beta$ represents any constant overhead of the compression -- encoding dictionaries, block boundaries, etc. -- that arises from the difference between the nearly incompressible random configuration and the simulated configuration. The scaling factor $\alpha \cong 0.425$ represents imperfections in the compression algorithm and any other factors -- e.g., a factor of $\ln 2$ due to choice of base. The linear relation holds in the region where there is most interest in these systems, i.e., in the dense liquid region corresponding to values of excess entropy lower than about $-1$. While the focus of this paper has been on such typical liquid state points, we note that the entropy of solids should also be straightforward to estimate from optimal data compression.

To summarize, we have shown that the diffusion coefficient of simple liquids can be estimated reliably from the optimal compression of a single equilibrium configuration as a proxy for the excess entropy. This should also work for extremely viscous liquids, which can only be equilibrated by SWAP methods and for which thermodynamic integration therefore is not an option to determine the excess entropy. If our findings can be extended to realistic molecular models, estimating dynamic properties of liquids like the diffusion coefficient or viscosity will become much easier than at present.

\begin{acknowledgments}
We thank Daniele Coslovich for useful comments on the  manuscript.
This work was supported by the VILLUM Foundation's \textit{Matter} grant VIL16515.
\end{acknowledgments}

\section*{Supplemental Material}
Details on the compression methods used are given here.

\subsection*{Compression methods}

The compression methods used are the following:

\begin{description}[style=nextline]
	
	\item[Brotli LEB128] An integer quantization which is turned into a variable-size integer via base128 encoding (as opposed to a fixed 16/32/etc. bit number), then fed through the Brotli (a DEFLATE/LZ77+Huffman derivative) compressor \cite{alakuijala2018brotli,brotliGithub}. We use a quantization of 1/1024 here, keeping enough of the precision to accurately capture the important bits of the configuration, but leaving out the less structurally important very-short time details. The aim is to be able to reconstruct a configuration that looks ``close enough'' to the original one.
	
	\item[brotliF32] Instead of an integer quantization, here we use the 32-bit floating-point number generated from our previous transformations, although we now scale the radial distance by the cube root of the density of the system in order to normalize the value --- a concession to the nature of floating point data. While this does change the effective quantization, we are now including more bits of data in the compression, far exceeding the resolution of the above quantization. The compressed size is much larger as expected as it now contains many more bits of ``insignificant'' data, but the difference shows the same general shape and even has a larger $\Kex$.
	
	\item[zpaqF32] We also include the same 32-bit floating point representation, but using the zpaq compressor instead \cite{mahoney2009incremental,zpaqGithub}. This is a much slower compressor, which includes a variety of methods automatically chosen to best fit the compressed data, resulting in a compression that is approximately 8\% better than Brotli.
	
	\item[zpaqF16] Finally, we also include a 16-bit floating point, zpaq compressed set, which is approximately the same resolution as the integer quantization.
\end{description}

\subsection*{Extended methodology}

The computation of $\Kex$ from simulation data proceeds as follows:

\begin{enumerate}
	\item A configuration is given as the absolute ($x,y,z$) Float32 coordinates within a (orthorhombic) simulation cell.
	\item An arbitrary particle is taken to be the origin and the cell boundaries translated such that it now lies at ($0,0,0$) with all particles kept within the bounds of the cell.
	\item To arrange the particles, a Hilbert curve of sufficient order to provide unique coordinates for every particle is created and used to sort the particles in a spatially ordered manner.
	\item The absolute ($x,y,z$) coordinates are transformed into ($\Delta x, \Delta y, \Delta z$) tuples by subtracting the absolute coordinates of the previous particle. We are thus left with $N-1$ ($\Delta x, \Delta y, \Delta z$) tuples.
	\item The ($\Delta x, \Delta y, \Delta z$) tuples are converted to ``internal'' coordinates  -- distance, angle, torsion -- ($r, \theta, \phi$) as used in inter-molecular coordinates. We now have $N-1$ distances, $N-2$ angles, and $N-3$ torsions; for simplicity we fill the missing angle and torsions of the first few particles with 1.0 to keep $N-1$ ($r, \theta, \phi$) tuples.
\end{enumerate}

This leaves $N-1$ ($r, \theta, \phi$) tuples (still in Float32 format) that we proceed to compress to estimate the Kolmogorov complexity $K$. From here, we can choose any pre-processing steps, e.g., to quantize the Float32 data to an integer, downcast to a Float16, byte reordering, etc. We found that it is usually optimal at least to ``de-interleave'' the tuples into their own $r, \theta,$ and $\phi$ vectors, as the similarities within the components otherwise were less pronounced. These data are then compressed, either by writing to disk (in raw binary) and calling a program, or for some methods via in-memory functions. The length of the resulting compressed representation (converted to bits) was then divided by $N$ to determine $K$.

For the ``ideal gas'' configuration, $N$ particles were randomly distributed in the same size simulation cell and the same procedure was carried out. $\Kex$ is estimated as the difference between optimal data compression of the as-simulated configuration and that of the random configuration. A sample from $\rho = 0.84, T = 2.0$ of the LJ system is shown in \Cref{tab:comprSize}, resulting in $\Kex = -1.9735$ bits/particle.

\begin{table}
	\centering
	\begin{tabular}{lrrrr}
		& $r$ & $\theta$ & $\phi$ & Sum\\
		Uncompressed            & 127\,996 & 127\,996 & 127\,996 & 383\,988 \\
		Equil. (Compressed)    & 94\,363 & 97\,833 & 101\,390 & 293\,586 \\
		Rand. (Compressed)      & 100\,619 & 99\,042 & 101\,819 & 301\,480 \\
		Delta                   & -6\,256 & -1\,209 & -429 & -7\,894 \\
	\end{tabular}
	\caption{Size in bytes of the various compression stages for a 32\,000 particle configuration from $\rho = 0.84, T = 2.0$ of the LJ system. Both the random and equilibrium configuration share the same uncompressed size.}
	\label{tab:comprSize}
\end{table}

A number of other compression methods have been tested, including floating-point specific programs. However, many of these specific programs either focus on throughput (e.g., for moving data between GPU/host or over networks) and/or smoother data than we have. None of them worked better than zpaqF32, and many failed to beat other general-purpose compressors with simple pre-processing of the data (e.g., byte reordering/shuffling).

\subsection*{Single Configuration}
\begin{figure}
	\centering
	\includegraphics[width=7cm]{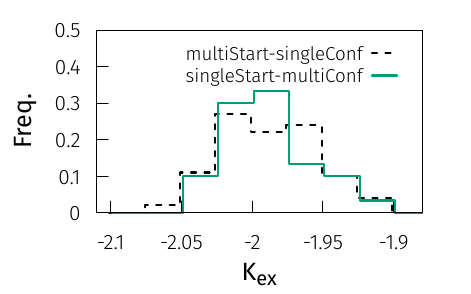}
	\caption{Comparison of $\Kex$ from 30 independent configurations sampled once with one configuration sampled 100 times. $T = 2.0, \rho = 0.84$}
	\label{fig:multistart}
\end{figure}

\Fig{fig:multistart} shows that re-sampling a single configuration using different initial particles results in the same distribution of $\Kex$ as sampling independent configurations once each.

\subsection*{Internal Coordinates}

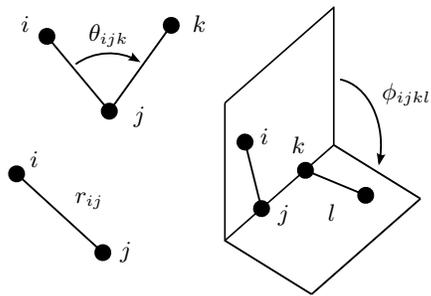
\begin{figure}
	\centering
	\begin{overpic}[width=6cm]{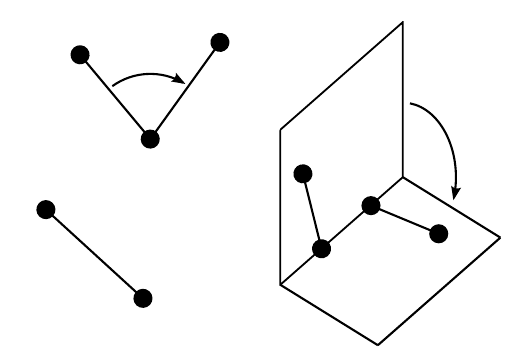}
		\put (12,30) {$i$}
		\put (32,10) {$j$}
		\put (22,22) {$r_{ij}$}
		
		\put (10,60) {$i$}
		\put (35,40) {$j$}
		\put (48,60) {$k$}
		\put (25,58) {$\theta_{ijk}$}
		
		\put (63,35) {$i$}
		\put (67,18) {$j$}
		\put (70,33) {$k$}
		\put (78,18) {$l$}
		\put (90,45) {$\phi_{ijkl}$}
	\end{overpic}
	\caption{Internal coordinate definitions using up to four particles $i,j,k,l$}
	\label{fig:iCoord}
\end{figure}

\Fig{fig:iCoord} illustrates the construction of $r, \theta, \phi$ used in the internal coordinate representation.
To define the position of particle $n$, we use particles $n, n-1, n-2, n-3$ as $i,j,k,l$ respectively.
For the first few particles, some of these values are undefined as they represent the (discarded) rotational degrees of freedom of the system.

\begin{thebibliography}{47}%
	\makeatletter
	\providecommand \@ifxundefined [1]{%
		\@ifx{#1\undefined}
	}%
	\providecommand \@ifnum [1]{%
		\ifnum #1\expandafter \@firstoftwo
		\else \expandafter \@secondoftwo
		\fi
	}%
	\providecommand \@ifx [1]{%
		\ifx #1\expandafter \@firstoftwo
		\else \expandafter \@secondoftwo
		\fi
	}%
	\providecommand \natexlab [1]{#1}%
	\providecommand \enquote  [1]{``#1''}%
	\providecommand \bibnamefont  [1]{#1}%
	\providecommand \bibfnamefont [1]{#1}%
	\providecommand \citenamefont [1]{#1}%
	\providecommand \href@noop [0]{\@secondoftwo}%
	\providecommand \href [0]{\begingroup \@sanitize@url \@href}%
	\providecommand \@href[1]{\@@startlink{#1}\@@href}%
	\providecommand \@@href[1]{\endgroup#1\@@endlink}%
	\providecommand \@sanitize@url [0]{\catcode `\\12\catcode `\$12\catcode
		`\&12\catcode `\#12\catcode `\^12\catcode `\_12\catcode `\%12\relax}%
	\providecommand \@@startlink[1]{}%
	\providecommand \@@endlink[0]{}%
	\providecommand \url  [0]{\begingroup\@sanitize@url \@url }%
	\providecommand \@url [1]{\endgroup\@href {#1}{\urlprefix }}%
	\providecommand \urlprefix  [0]{URL }%
	\providecommand \Eprint [0]{\href }%
	\providecommand \doibase [0]{https://doi.org/}%
	\providecommand \selectlanguage [0]{\@gobble}%
	\providecommand \bibinfo  [0]{\@secondoftwo}%
	\providecommand \bibfield  [0]{\@secondoftwo}%
	\providecommand \translation [1]{[#1]}%
	\providecommand \BibitemOpen [0]{}%
	\providecommand \bibitemStop [0]{}%
	\providecommand \bibitemNoStop [0]{.\EOS\space}%
	\providecommand \EOS [0]{\spacefactor3000\relax}%
	\providecommand \BibitemShut  [1]{\csname bibitem#1\endcsname}%
	\let\auto@bib@innerbib\@empty
	\bibitem [{\citenamefont {Rovelli}(2018)}]{rovelli}%
	\BibitemOpen
	\bibfield  {author} {\bibinfo {author} {\bibfnamefont {C.}~\bibnamefont
			{Rovelli}},\ }\href@noop {} {\emph {\bibinfo {title} {{The Order of Time}}}}\
	(\bibinfo  {publisher} {Penguin Books, London},\ \bibinfo {year} {2018})\BibitemShut
	{NoStop}%
	\bibitem [{\citenamefont {Landau}\ and\ \citenamefont
		{Lifshitz}(1958)}]{LLstat}%
	\BibitemOpen
	\bibfield  {author} {\bibinfo {author} {\bibfnamefont {L.~D.}\ \bibnamefont
			{Landau}}\ and\ \bibinfo {author} {\bibfnamefont {E.~M.}\ \bibnamefont
			{Lifshitz}},\ }\href@noop {} {\emph {\bibinfo {title} {Statistical
				Physics}}}\ (\bibinfo  {publisher} {Pergamon, Oxford},\ \bibinfo {year}
	{1958})\BibitemShut {NoStop}%
	\bibitem [{\citenamefont {Reichl}(2016)}]{reichl}%
	\BibitemOpen
	\bibfield  {author} {\bibinfo {author} {\bibfnamefont {L.~E.}\ \bibnamefont
			{Reichl}},\ }\href@noop {} {\emph {\bibinfo {title} {{A Modern Course in
					Statistical Physics}}}},\ \bibinfo {edition} {4th}\ ed.\ (\bibinfo
	{publisher} {Wiley-VCH},\ \bibinfo {year} {2016})\BibitemShut {NoStop}%
	\bibitem [{\citenamefont {Rosenfeld}(1977)}]{ros77}%
	\BibitemOpen
	\bibfield  {author} {\bibinfo {author} {\bibfnamefont {Y.}~\bibnamefont
			{Rosenfeld}},\ }\bibfield  {title} {\bibinfo {title} {Relation between the
			transport coefficients and the internal entropy of simple systems},\
	}\href@noop {} {\bibfield  {journal} {\bibinfo  {journal} {Phys. Rev. A}\
		}\textbf {\bibinfo {volume} {15}},\ \bibinfo {pages} {2545} (\bibinfo {year}
		{1977})}\BibitemShut {NoStop}%
	\bibitem [{\citenamefont {Dyre}(2018)}]{dyr18a}%
	\BibitemOpen
	\bibfield  {author} {\bibinfo {author} {\bibfnamefont {J.~C.}\ \bibnamefont
			{Dyre}},\ }\bibfield  {title} {\bibinfo {title} {Perspective: Excess-entropy
			scaling},\ }\href {https://doi.org/10.1063/1.5055064} {\bibfield  {journal}
		{\bibinfo  {journal} {J. Chem. Phys.}\ }\textbf {\bibinfo {volume} {149}},\
		\bibinfo {pages} {210901} (\bibinfo {year} {2018})}\BibitemShut {NoStop}%
	\bibitem [{\citenamefont {Novak}(2013)}]{nov13}%
	\BibitemOpen
	\bibfield  {author} {\bibinfo {author} {\bibfnamefont {L.~T.}\ \bibnamefont
			{Novak}},\ }\bibfield  {title} {\bibinfo {title} {Predictive
			corresponding-states viscosity model for the entire fluid region:
			n-alkanes},\ }\href {https://doi.org/10.1021/ie400654p} {\bibfield  {journal}
		{\bibinfo  {journal} {Ind. Eng. Chem. Res.}\ }\textbf {\bibinfo {volume}
			{52}},\ \bibinfo {pages} {6841} (\bibinfo {year} {2013})}\BibitemShut
	{NoStop}%
	\bibitem [{\citenamefont {Hopp}\ and\ \citenamefont {Gross}(2017)}]{hop17}%
	\BibitemOpen
	\bibfield  {author} {\bibinfo {author} {\bibfnamefont {M.}~\bibnamefont
			{Hopp}}\ and\ \bibinfo {author} {\bibfnamefont {J.}~\bibnamefont {Gross}},\
	}\bibfield  {title} {\bibinfo {title} {Thermal conductivity of real
			substances from excess entropy scaling using {PCP-SAFT}},\ }\href
	{https://doi.org/10.1021/acs.iecr.6b04289} {\bibfield  {journal} {\bibinfo
			{journal} {Ind. Eng. Chem. Res.}\ }\textbf {\bibinfo {volume} {56}},\
		\bibinfo {pages} {4527} (\bibinfo {year} {2017})}\BibitemShut {NoStop}%
	\bibitem [{\citenamefont {Fouad}\ and\ \citenamefont {Vega}(2018)}]{fou18}%
	\BibitemOpen
	\bibfield  {author} {\bibinfo {author} {\bibfnamefont {W.~A.}\ \bibnamefont
			{Fouad}}\ and\ \bibinfo {author} {\bibfnamefont {L.~F.}\ \bibnamefont
			{Vega}},\ }\bibfield  {title} {\bibinfo {title} {Transport properties of
			{HFC} and {HFO} based refrigerants using an excess entropy scaling
			approach},\ }\href
	{https://doi.org/https://doi.org/10.1016/j.supflu.2017.09.006} {\bibfield
		{journal} {\bibinfo  {journal} {J. Supercrit. Fluids}\ }\textbf {\bibinfo
			{volume} {131}},\ \bibinfo {pages} {106 } (\bibinfo {year}
		{2018})}\BibitemShut {NoStop}%
	\bibitem [{\citenamefont {Bastea}(2011)}]{bas11}%
	\BibitemOpen
	\bibfield  {author} {\bibinfo {author} {\bibfnamefont {S.}~\bibnamefont
			{Bastea}},\ }\bibfield  {title} {\bibinfo {title} {Thermodynamics and
			diffusion in size-symmetric and asymmetric dense electrolytes},\ }\href
	{https://doi.org/10.1063/1.3629782} {\bibfield  {journal} {\bibinfo
			{journal} {J. Chem. Phys.}\ }\textbf {\bibinfo {volume} {135}},\ \bibinfo
		{pages} {084515} (\bibinfo {year} {2011})}\BibitemShut {NoStop}%
	\bibitem [{\citenamefont {Goel}\ \emph {et~al.}(2011)\citenamefont {Goel},
		\citenamefont {Lacks},\ and\ \citenamefont {Van~Orman}}]{goe11}%
	\BibitemOpen
	\bibfield  {author} {\bibinfo {author} {\bibfnamefont {G.}~\bibnamefont
			{Goel}}, \bibinfo {author} {\bibfnamefont {D.~J.}\ \bibnamefont {Lacks}},\
		and\ \bibinfo {author} {\bibfnamefont {J.~A.}\ \bibnamefont {Van~Orman}},\
	}\bibfield  {title} {\bibinfo {title} {Transport coefficients in silicate
			melts from structural data via a structure-thermodynamics-dynamics
			relationship},\ }\href {https://doi.org/10.1103/PhysRevE.84.051506}
	{\bibfield  {journal} {\bibinfo  {journal} {Phys. Rev. E}\ }\textbf {\bibinfo
			{volume} {84}},\ \bibinfo {pages} {051506} (\bibinfo {year}
		{2011})}\BibitemShut {NoStop}%
	\bibitem [{\citenamefont {Jia}\ \emph {et~al.}(2015)\citenamefont {Jia},
		\citenamefont {Yun},\ and\ \citenamefont {Jianzhong}}]{jia15}%
	\BibitemOpen
	\bibfield  {author} {\bibinfo {author} {\bibfnamefont {F.}~\bibnamefont
			{Jia}}, \bibinfo {author} {\bibfnamefont {T.}~\bibnamefont {Yun}},\ and\
		\bibinfo {author} {\bibfnamefont {W.}~\bibnamefont {Jianzhong}},\ }\bibfield
	{title} {\bibinfo {title} {Classical density functional theory for methane
			adsorption in metal-organic framework materials},\ }\href
	{https://doi.org/10.1002/aic.14877} {\bibfield  {journal} {\bibinfo
			{journal} {AIChE Journal}\ }\textbf {\bibinfo {volume} {61}},\ \bibinfo
		{pages} {3012} (\bibinfo {year} {2015})}\BibitemShut {NoStop}%
	\bibitem [{\citenamefont {Liu}\ \emph {et~al.}(2016)\citenamefont {Liu},
		\citenamefont {Guo}, \citenamefont {Hu}, \citenamefont {Zhao}, \citenamefont
		{Liu},\ and\ \citenamefont {Hu}}]{liu16}%
	\BibitemOpen
	\bibfield  {author} {\bibinfo {author} {\bibfnamefont {Y.}~\bibnamefont
			{Liu}}, \bibinfo {author} {\bibfnamefont {F.}~\bibnamefont {Guo}}, \bibinfo
		{author} {\bibfnamefont {J.}~\bibnamefont {Hu}}, \bibinfo {author}
		{\bibfnamefont {S.}~\bibnamefont {Zhao}}, \bibinfo {author} {\bibfnamefont
			{H.}~\bibnamefont {Liu}},\ and\ \bibinfo {author} {\bibfnamefont
			{Y.}~\bibnamefont {Hu}},\ }\bibfield  {title} {\bibinfo {title} {Entropy
			prediction for {${\rm H_2}$} adsorption in metal-organic frameworks},\ }\href
	{https://doi.org/10.1039/C6CP04645B} {\bibfield  {journal} {\bibinfo
			{journal} {Phys. Chem. Chem. Phys.}\ }\textbf {\bibinfo {volume} {18}},\
		\bibinfo {pages} {23998} (\bibinfo {year} {2016})}\BibitemShut {NoStop}%
	\bibitem [{\citenamefont {Cao}\ \emph {et~al.}(2016)\citenamefont {Cao},
		\citenamefont {Wang}, \citenamefont {Shao},\ and\ \citenamefont
		{Wang}}]{cao16}%
	\BibitemOpen
	\bibfield  {author} {\bibinfo {author} {\bibfnamefont {Q.-L.}\ \bibnamefont
			{Cao}}, \bibinfo {author} {\bibfnamefont {P.-P.}\ \bibnamefont {Wang}},
		\bibinfo {author} {\bibfnamefont {J.-X.}\ \bibnamefont {Shao}},\ and\
		\bibinfo {author} {\bibfnamefont {F.-H.}\ \bibnamefont {Wang}},\ }\bibfield
	{title} {\bibinfo {title} {Transport properties and entropy-scaling laws for
			diffusion coefficients in liquid {${\rm Fe_{0.9}Ni_{0.1}}$} up to 350
			{GPa}},\ }\href {https://doi.org/10.1039/C6RA15211B} {\bibfield  {journal}
		{\bibinfo  {journal} {RSC Adv.}\ }\textbf {\bibinfo {volume} {6}},\ \bibinfo
		{pages} {84420} (\bibinfo {year} {2016})}\BibitemShut {NoStop}%
	\bibitem [{\citenamefont {Liu}\ \emph {et~al.}(2013)\citenamefont {Liu},
		\citenamefont {Fu},\ and\ \citenamefont {Wu}}]{liu13}%
	\BibitemOpen
	\bibfield  {author} {\bibinfo {author} {\bibfnamefont {Y.}~\bibnamefont
			{Liu}}, \bibinfo {author} {\bibfnamefont {J.}~\bibnamefont {Fu}},\ and\
		\bibinfo {author} {\bibfnamefont {J.}~\bibnamefont {Wu}},\ }\bibfield
	{title} {\bibinfo {title} {Excess-entropy scaling for gas diffusivity in
			nanoporous materials},\ }\href {https://doi.org/10.1021/la403082q} {\bibfield
		{journal} {\bibinfo  {journal} {Langmuir}\ }\textbf {\bibinfo {volume}
			{29}},\ \bibinfo {pages} {12997} (\bibinfo {year} {2013})}\BibitemShut
	{NoStop}%
	\bibitem [{\citenamefont {Tian}\ \emph {et~al.}(2018)\citenamefont {Tian},
		\citenamefont {Fei},\ and\ \citenamefont {Wu}}]{tia18}%
	\BibitemOpen
	\bibfield  {author} {\bibinfo {author} {\bibfnamefont {Y.}~\bibnamefont
			{Tian}}, \bibinfo {author} {\bibfnamefont {W.}~\bibnamefont {Fei}},\ and\
		\bibinfo {author} {\bibfnamefont {J.}~\bibnamefont {Wu}},\ }\bibfield
	{title} {\bibinfo {title} {Separation of carbon isotopes in methane with
			nanoporous materials},\ }\href {https://doi.org/10.1021/acs.iecr.8b00364}
	{\bibfield  {journal} {\bibinfo  {journal} {Ind. Eng. Chem. Res.}\ }\textbf
		{\bibinfo {volume} {57}},\ \bibinfo {pages} {5151} (\bibinfo {year}
		{2018})}\BibitemShut {NoStop}%
	\bibitem [{\citenamefont {Gnan}\ \emph {et~al.}(2009)\citenamefont {Gnan},
		\citenamefont {Schr{\o}der}, \citenamefont {Pedersen}, \citenamefont
		{Bailey},\ and\ \citenamefont {Dyre}}]{IV}%
	\BibitemOpen
	\bibfield  {author} {\bibinfo {author} {\bibfnamefont {N.}~\bibnamefont
			{Gnan}}, \bibinfo {author} {\bibfnamefont {T.~B.}\ \bibnamefont
			{Schr{\o}der}}, \bibinfo {author} {\bibfnamefont {U.~R.}\ \bibnamefont
			{Pedersen}}, \bibinfo {author} {\bibfnamefont {N.~P.}\ \bibnamefont
			{Bailey}},\ and\ \bibinfo {author} {\bibfnamefont {J.~C.}\ \bibnamefont
			{Dyre}},\ }\bibfield  {title} {\bibinfo {title} {Pressure-energy correlations
			in liquids. {IV. ``Isomorphs''} in liquid phase diagrams},\ }\href
	{https://doi.org/10.1063/1.3265957} {\bibfield  {journal} {\bibinfo
			{journal} {J. Chem. Phys.}\ }\textbf {\bibinfo {volume} {131}},\ \bibinfo
		{pages} {234504} (\bibinfo {year} {2009})}\BibitemShut {NoStop}%
	\bibitem [{\citenamefont {Dyre}(2014)}]{dyr14}%
	\BibitemOpen
	\bibfield  {author} {\bibinfo {author} {\bibfnamefont {J.~C.}\ \bibnamefont
			{Dyre}},\ }\bibfield  {title} {\bibinfo {title} {Hidden scale invariance in
			condensed matter},\ }\href@noop {} {\bibfield  {journal} {\bibinfo  {journal}
			{J. Phys. Chem. B}\ }\textbf {\bibinfo {volume} {118}},\ \bibinfo {pages}
		{10007} (\bibinfo {year} {2014})}\BibitemShut {NoStop}%
	\bibitem [{\citenamefont {Schr{\o}der}\ and\ \citenamefont
		{Dyre}(2014)}]{sch14}%
	\BibitemOpen
	\bibfield  {author} {\bibinfo {author} {\bibfnamefont {T.~B.}\ \bibnamefont
			{Schr{\o}der}}\ and\ \bibinfo {author} {\bibfnamefont {J.~C.}\ \bibnamefont
			{Dyre}},\ }\bibfield  {title} {\bibinfo {title} {Simplicity of condensed
			matter at its core: Generic definition of a {Roskilde}-simple system},\
	}\href {https://doi.org/http://dx.doi.org/10.1063/1.4901215} {\bibfield
		{journal} {\bibinfo  {journal} {J. Chem. Phys.}\ }\textbf {\bibinfo {volume}
			{141}},\ \bibinfo {pages} {204502} (\bibinfo {year} {2014})}\BibitemShut
	{NoStop}%
	\bibitem [{\citenamefont {Bell}(2019)}]{Bel19}%
	\BibitemOpen
	\bibfield  {author} {\bibinfo {author} {\bibfnamefont {I.~H.}\ \bibnamefont
			{Bell}},\ }\bibfield  {title} {\bibinfo {title} {Probing the link between
			residual entropy and viscosity of molecular fluids and model potentials},\
	}\href {https://doi.org/10.1073/pnas.1815943116} {\bibfield  {journal}
		{\bibinfo  {journal} {Proc. Nat. Acad. Sci. (USA)}\ }\textbf {\bibinfo
			{volume} {116}},\ \bibinfo {pages} {4070} (\bibinfo {year}
		{2019})}\BibitemShut {NoStop}%
	\bibitem [{\citenamefont {Bell}\ \emph {et~al.}(2019)\citenamefont {Bell},
		\citenamefont {Messerly}, \citenamefont {Thol}, \citenamefont {Costigliola},\
		and\ \citenamefont {Dyre}}]{bel19a}%
	\BibitemOpen
	\bibfield  {author} {\bibinfo {author} {\bibfnamefont {I.~H.}\ \bibnamefont
			{Bell}}, \bibinfo {author} {\bibfnamefont {R.}~\bibnamefont {Messerly}},
		\bibinfo {author} {\bibfnamefont {M.}~\bibnamefont {Thol}}, \bibinfo {author}
		{\bibfnamefont {L.}~\bibnamefont {Costigliola}},\ and\ \bibinfo {author}
		{\bibfnamefont {J.~C.}\ \bibnamefont {Dyre}},\ }\bibfield  {title} {\bibinfo
		{title} {Modified entropy scaling of the transport properties of the
			{Lennard-Jones} fluid},\ }\href {https://doi.org/10.1021/acs.jpcb.9b05808}
	{\bibfield  {journal} {\bibinfo  {journal} {J. Phys. Chem. B}\ }\textbf
		{\bibinfo {volume} {123}},\ \bibinfo {pages} {6345} (\bibinfo {year}
		{2019})}\BibitemShut {NoStop}%
	\bibitem [{\citenamefont {Fermi}(1937)}]{Fermi1937thermo}%
	\BibitemOpen
	\bibfield  {author} {\bibinfo {author} {\bibfnamefont {E.}~\bibnamefont
			{Fermi}},\ }\href@noop {} {\emph {\bibinfo {title} {Thermodynamics}}},\ Dover
	Publications\ (\bibinfo  {publisher} {Prentice-Hall},\ \bibinfo {year}
	{1937})\BibitemShut {NoStop}%
	\bibitem [{\citenamefont {Frenkel}\ and\ \citenamefont {Smit}(2002)}]{frenkel}%
	\BibitemOpen
	\bibfield  {author} {\bibinfo {author} {\bibfnamefont {D.}~\bibnamefont
			{Frenkel}}\ and\ \bibinfo {author} {\bibfnamefont {B.}~\bibnamefont {Smit}},\
	}\href@noop {} {\emph {\bibinfo {title} {Understanding Molecular
				Simulation}}}\ (\bibinfo  {publisher} {Academic Press},\ \bibinfo {year}
	{2002})\BibitemShut {NoStop}%
	\bibitem [{\citenamefont {Hansen}\ and\ \citenamefont
		{McDonald}(2013)}]{han13}%
	\BibitemOpen
	\bibfield  {author} {\bibinfo {author} {\bibfnamefont {J.-P.}\ \bibnamefont
			{Hansen}}\ and\ \bibinfo {author} {\bibfnamefont {I.~R.}\ \bibnamefont
			{McDonald}},\ }\href@noop {} {\emph {\bibinfo {title} {{Theory of Simple
					Liquids: With Applications to Soft Matter}}}},\ \bibinfo {edition} {4th}\
	ed.\ (\bibinfo  {publisher} {Academic, New York},\ \bibinfo {year}
	{2013})\BibitemShut {NoStop}%
	\bibitem [{\citenamefont {Nettleton}\ and\ \citenamefont
		{Green}(1958)}]{net58}%
	\BibitemOpen
	\bibfield  {author} {\bibinfo {author} {\bibfnamefont {R.~E.}\ \bibnamefont
			{Nettleton}}\ and\ \bibinfo {author} {\bibfnamefont {M.~S.}\ \bibnamefont
			{Green}},\ }\bibfield  {title} {\bibinfo {title} {Expression in terms of
			molecular distribution functions for the entropy density in an infinite
			system},\ }\href@noop {} {\bibfield  {journal} {\bibinfo  {journal} {J. Chem.
				Phys.}\ }\textbf {\bibinfo {volume} {29}},\ \bibinfo {pages} {1365} (\bibinfo
		{year} {1958})}\BibitemShut {NoStop}%
	\bibitem [{\citenamefont {Baranyai}\ and\ \citenamefont {Evans}(1989)}]{bar89}%
	\BibitemOpen
	\bibfield  {author} {\bibinfo {author} {\bibfnamefont {A.}~\bibnamefont
			{Baranyai}}\ and\ \bibinfo {author} {\bibfnamefont {D.~J.}\ \bibnamefont
			{Evans}},\ }\bibfield  {title} {\bibinfo {title} {Direct entropy calculation
			from computer simulation of liquids},\ }\href
	{https://doi.org/10.1103/PhysRevA.40.3817} {\bibfield  {journal} {\bibinfo
			{journal} {Phys. Rev. A}\ }\textbf {\bibinfo {volume} {40}},\ \bibinfo
		{pages} {3817} (\bibinfo {year} {1989})}\BibitemShut {NoStop}%
	\bibitem [{\citenamefont {Green}(1952)}]{green52}%
	\BibitemOpen
	\bibfield  {author} {\bibinfo {author} {\bibfnamefont {H.~S.}\ \bibnamefont
			{Green}},\ }\href@noop {} {\emph {\bibinfo {title} {The Molecular Theory of
				Fluids}}}\ (\bibinfo  {publisher} {North-Holland, Amsterdam},\ \bibinfo
	{year} {1952})\BibitemShut {NoStop}%
	\bibitem [{\citenamefont {Widom}(1963)}]{wid63}%
	\BibitemOpen
	\bibfield  {author} {\bibinfo {author} {\bibfnamefont {B.}~\bibnamefont
			{Widom}},\ }\bibfield  {title} {\bibinfo {title} {Some topics in the theory
			of fluids},\ }\href {https://doi.org/10.1063/1.1734110} {\bibfield  {journal}
		{\bibinfo  {journal} {J. Chem. Phys.}\ }\textbf {\bibinfo {volume} {39}},\
		\bibinfo {pages} {2808} (\bibinfo {year} {1963})}\BibitemShut {NoStop}%
	\bibitem [{\citenamefont {Avinery}\ \emph {et~al.}(2019)\citenamefont
		{Avinery}, \citenamefont {Kornreich},\ and\ \citenamefont {Beck}}]{avi19}%
	\BibitemOpen
	\bibfield  {author} {\bibinfo {author} {\bibfnamefont {R.}~\bibnamefont
			{Avinery}}, \bibinfo {author} {\bibfnamefont {M.}~\bibnamefont {Kornreich}},\
		and\ \bibinfo {author} {\bibfnamefont {R.}~\bibnamefont {Beck}},\ }\bibfield
	{title} {\bibinfo {title} {Universal and accessible entropy estimation using
			a compression algorithm},\ }\href
	{https://doi.org/10.1103/PhysRevLett.123.178102} {\bibfield  {journal}
		{\bibinfo  {journal} {Phys. Rev. Lett.}\ }\textbf {\bibinfo {volume} {123}},\
		\bibinfo {pages} {178102} (\bibinfo {year} {2019})}\BibitemShut {NoStop}%
	\bibitem [{\citenamefont {Martiniani}\ \emph {et~al.}(2019)\citenamefont
		{Martiniani}, \citenamefont {Chaikin},\ and\ \citenamefont {Levine}}]{mar19}%
	\BibitemOpen
	\bibfield  {author} {\bibinfo {author} {\bibfnamefont {S.}~\bibnamefont
			{Martiniani}}, \bibinfo {author} {\bibfnamefont {P.~M.}\ \bibnamefont
			{Chaikin}},\ and\ \bibinfo {author} {\bibfnamefont {D.}~\bibnamefont
			{Levine}},\ }\bibfield  {title} {\bibinfo {title} {Quantifying hidden order
			out of equilibrium},\ }\href {https://doi.org/10.1103/PhysRevX.9.011031}
	{\bibfield  {journal} {\bibinfo  {journal} {Phys. Rev. X}\ }\textbf {\bibinfo
			{volume} {9}},\ \bibinfo {pages} {011031} (\bibinfo {year}
		{2019})}\BibitemShut {NoStop}%
	\bibitem [{\citenamefont {Cover}\ and\ \citenamefont
		{Thomas}(2006)}]{Cover_and_Thomas}%
	\BibitemOpen
	\bibfield  {author} {\bibinfo {author} {\bibfnamefont {T.~M.}\ \bibnamefont
			{Cover}}\ and\ \bibinfo {author} {\bibfnamefont {J.~A.}\ \bibnamefont
			{Thomas}},\ }\href@noop {} {\emph {\bibinfo {title} {{Elements of Information
					Theory}}}},\ \bibinfo {edition} {2nd}\ ed.\ (\bibinfo  {publisher} {Wiley},\
	\bibinfo {year} {2006})\BibitemShut {NoStop}%
	\bibitem [{\citenamefont {Li}\ and\ \citenamefont
		{Vitanyi}(2008)}]{Li_and_Vitanyi}%
	\BibitemOpen
	\bibfield  {author} {\bibinfo {author} {\bibfnamefont {M.}~\bibnamefont
			{Li}}\ and\ \bibinfo {author} {\bibfnamefont {P.}~\bibnamefont {Vitanyi}},\
	}\href@noop {} {\emph {\bibinfo {title} {{An Introduction to Kolmogorov
					Complexity and Its Applications}}}},\ \bibinfo {edition} {3rd}\ ed.\
	(\bibinfo  {publisher} {Springer},\ \bibinfo {year} {2008})\BibitemShut
	{NoStop}%
	\bibitem [{\citenamefont {Lennard-Jones}(1924)}]{lj24}%
	\BibitemOpen
	\bibfield  {author} {\bibinfo {author} {\bibfnamefont {J.~E.}\ \bibnamefont
			{Lennard-Jones}},\ }\bibfield  {title} {\bibinfo {title} {On the
			determination of molecular fields. {I. From} the variation of the viscosity
			of a gas with temperature},\ }\href@noop {} {\bibfield  {journal} {\bibinfo
			{journal} {Proc. R. Soc. London A}\ }\textbf {\bibinfo {volume} {106}},\
		\bibinfo {pages} {441} (\bibinfo {year} {1924})}\BibitemShut {NoStop}%
	\bibitem [{\citenamefont {Morse}(1929)}]{Morse1929}%
	\BibitemOpen
	\bibfield  {author} {\bibinfo {author} {\bibfnamefont {P.~M.}\ \bibnamefont
			{Morse}},\ }\bibfield  {title} {\bibinfo {title} {Diatomic molecules
			according to the wave mechanics. {II}. {Vibrational} levels},\ }\href
	{https://doi.org/10.1103/PhysRev.34.57} {\bibfield  {journal} {\bibinfo
			{journal} {Phys. Rev.}\ }\textbf {\bibinfo {volume} {34}},\ \bibinfo {pages}
		{57} (\bibinfo {year} {1929})}\BibitemShut {NoStop}%
	\bibitem [{\citenamefont {Yukawa}(1935)}]{yuk35}%
	\BibitemOpen
	\bibfield  {author} {\bibinfo {author} {\bibfnamefont {H.}~\bibnamefont
			{Yukawa}},\ }\bibfield  {title} {\bibinfo {title} {On the interaction of
			elementary particles. {I}},\ }\href
	{https://doi.org/10.11429/ppmsj1919.17.0_48} {\bibfield  {journal} {\bibinfo
			{journal} {Proc. Phys. Math. Soc. Jpn.}\ }\textbf {\bibinfo {volume} {17}},\
		\bibinfo {pages} {48} (\bibinfo {year} {1935})}\BibitemShut {NoStop}%
	\bibitem [{\citenamefont {Bailey}\ \emph {et~al.}(2017)\citenamefont {Bailey},
		\citenamefont {Ingebrigtsen}, \citenamefont {Hansen}, \citenamefont
		{Veldhorst}, \citenamefont {B{\o}hling}, \citenamefont {Lemarchand},
		\citenamefont {Olsen}, \citenamefont {Bacher}, \citenamefont {Costigliola},
		\citenamefont {Pedersen}, \citenamefont {Larsen}, \citenamefont {Dyre},\ and\
		\citenamefont {Schr{\o}der}}]{RUMD}%
	\BibitemOpen
	\bibfield  {author} {\bibinfo {author} {\bibfnamefont {N.~P.}\ \bibnamefont
			{Bailey}}, \bibinfo {author} {\bibfnamefont {T.~S.}\ \bibnamefont
			{Ingebrigtsen}}, \bibinfo {author} {\bibfnamefont {J.~S.}\ \bibnamefont
			{Hansen}}, \bibinfo {author} {\bibfnamefont {A.~A.}\ \bibnamefont
			{Veldhorst}}, \bibinfo {author} {\bibfnamefont {L.}~\bibnamefont
			{B{\o}hling}}, \bibinfo {author} {\bibfnamefont {C.~A.}\ \bibnamefont
			{Lemarchand}}, \bibinfo {author} {\bibfnamefont {A.~E.}\ \bibnamefont
			{Olsen}}, \bibinfo {author} {\bibfnamefont {A.~K.}\ \bibnamefont {Bacher}},
		\bibinfo {author} {\bibfnamefont {L.}~\bibnamefont {Costigliola}}, \bibinfo
		{author} {\bibfnamefont {U.~R.}\ \bibnamefont {Pedersen}}, \bibinfo {author}
		{\bibfnamefont {H.}~\bibnamefont {Larsen}}, \bibinfo {author} {\bibfnamefont
			{J.~C.}\ \bibnamefont {Dyre}},\ and\ \bibinfo {author} {\bibfnamefont
			{T.~B.}\ \bibnamefont {Schr{\o}der}},\ }\bibfield  {title} {\bibinfo {title}
		{{RUMD}: A general purpose molecular dynamics package optimized to utilize
			{GPU} hardware down to a few thousand particles},\ }\href
	{https://doi.org/10.21468/SciPostPhys.3.6.038} {\bibfield  {journal}
		{\bibinfo  {journal} {Scipost Phys.}\ }\textbf {\bibinfo {volume} {3}},\
		\bibinfo {pages} {038} (\bibinfo {year} {2017})}\BibitemShut {NoStop}%
	\bibitem [{\citenamefont {Hilbert}(1891)}]{Hilbert}%
	\BibitemOpen
	\bibfield  {author} {\bibinfo {author} {\bibfnamefont {D.}~\bibnamefont
			{Hilbert}},\ }\bibfield  {title} {\bibinfo {title} {{{\"U}ber die stetige
				Abbildung einer Line auf ein Fl{\"a}chenst{\"u}ck}},\ }\href
	{https://www.digizeitschriften.de/id/235181684_0038|log40} {\bibfield
		{journal} {\bibinfo  {journal} {Mathematische Annalen}\ } (\bibinfo {year}
		{1891})}\BibitemShut {NoStop}%
	\bibitem [{\citenamefont {Walraven}\ and\ \citenamefont
		{Leermakers}(2020)}]{walraven2020entropy}%
	\BibitemOpen
	\bibfield  {author} {\bibinfo {author} {\bibfnamefont {E.}~\bibnamefont
			{Walraven}}\ and\ \bibinfo {author} {\bibfnamefont {F.}~\bibnamefont
			{Leermakers}},\ }\bibfield  {title} {\bibinfo {title} {Entropy estimates of a
			hard sphere system by data compression of {Monte Carlo} simulation data},\
	}\href@noop {} {\bibfield  {journal} {\bibinfo  {journal} {Soft Matter}\
		}\textbf {\bibinfo {volume} {16}},\ \bibinfo {pages} {3740} (\bibinfo {year}
		{2020})}\BibitemShut {NoStop}%
	\bibitem [{\citenamefont {Zu}\ \emph {et~al.}(2020)\citenamefont {Zu},
		\citenamefont {Bupathy}, \citenamefont {Frenkel},\ and\ \citenamefont
		{Sastry}}]{zu2020information}%
	\BibitemOpen
	\bibfield  {author} {\bibinfo {author} {\bibfnamefont {M.}~\bibnamefont
			{Zu}}, \bibinfo {author} {\bibfnamefont {A.}~\bibnamefont {Bupathy}},
		\bibinfo {author} {\bibfnamefont {D.}~\bibnamefont {Frenkel}},\ and\ \bibinfo
		{author} {\bibfnamefont {S.}~\bibnamefont {Sastry}},\ }\bibfield  {title}
	{\bibinfo {title} {Information density, structure and entropy in equilibrium
			and non-equilibrium systems},\ }\href@noop {} {\bibfield  {journal} {\bibinfo
			{journal} {Journal of Statistical Mechanics: Theory and Experiment}\
		}\textbf {\bibinfo {volume} {2020}},\ \bibinfo {pages} {023204} (\bibinfo
		{year} {2020})}\BibitemShut {NoStop}%
	\bibitem [{\citenamefont {Alakuijala}\ \emph {et~al.}(2018)\citenamefont
		{Alakuijala}, \citenamefont {Farruggia}, \citenamefont {Ferragina},
		\citenamefont {Kliuchnikov}, \citenamefont {Obryk}, \citenamefont
		{Szabadka},\ and\ \citenamefont {Vandevenne}}]{alakuijala2018brotli}%
	\BibitemOpen
	\bibfield  {author} {\bibinfo {author} {\bibfnamefont {J.}~\bibnamefont
			{Alakuijala}}, \bibinfo {author} {\bibfnamefont {A.}~\bibnamefont
			{Farruggia}}, \bibinfo {author} {\bibfnamefont {P.}~\bibnamefont
			{Ferragina}}, \bibinfo {author} {\bibfnamefont {E.}~\bibnamefont
			{Kliuchnikov}}, \bibinfo {author} {\bibfnamefont {R.}~\bibnamefont {Obryk}},
		\bibinfo {author} {\bibfnamefont {Z.}~\bibnamefont {Szabadka}},\ and\
		\bibinfo {author} {\bibfnamefont {L.}~\bibnamefont {Vandevenne}},\ }\bibfield
	{title} {\bibinfo {title} {Brotli: A general-purpose data compressor},\
	}\href@noop {} {\bibfield  {journal} {\bibinfo  {journal} {ACM Transactions
				on Information Systems (TOIS)}\ }\textbf {\bibinfo {volume} {37}},\ \bibinfo
		{pages} {1} (\bibinfo {year} {2018})}\BibitemShut {NoStop}%
	\bibitem [{\citenamefont {Google}(2020)}]{brotliGithub}%
	\BibitemOpen
	\bibfield  {author} {\bibinfo {author} {\bibnamefont {Google}},\ }\href@noop
	{} {\bibinfo {title} {Brotli}},\ \bibinfo {howpublished}
	{\url{https://github.com/google/brotli}} (\bibinfo {year} {2020})\BibitemShut
	{NoStop}%
	\bibitem [{\citenamefont {Mahoney}(2009)}]{mahoney2009incremental}%
	\BibitemOpen
	\bibfield  {author} {\bibinfo {author} {\bibfnamefont {M.}~\bibnamefont
			{Mahoney}},\ }\href@noop {} {\bibinfo {title} {Incremental journaling backup
			utility and archiver}},\ \bibinfo {howpublished}
	{\url{https://mattmahoney.net/dc/zpaq.html}} (\bibinfo {year}
	{2009})\BibitemShut {NoStop}%
	\bibitem [{\citenamefont {Mahoney}(2016)}]{zpaqGithub}%
	\BibitemOpen
	\bibfield  {author} {\bibinfo {author} {\bibfnamefont {M.}~\bibnamefont
			{Mahoney}},\ }\href@noop {} {\bibinfo {title} {zpaq}},\ \bibinfo
	{howpublished} {\url{https://github.com/zpaq/zpaq}} (\bibinfo {year}
	{2016})\BibitemShut {NoStop}%
	\bibitem [{\citenamefont {Veldhorst}\ \emph {et~al.}(2015)\citenamefont
		{Veldhorst}, \citenamefont {Schr{\o}der},\ and\ \citenamefont
		{Dyre}}]{vel15}%
	\BibitemOpen
	\bibfield  {author} {\bibinfo {author} {\bibfnamefont {A.~A.}\ \bibnamefont
			{Veldhorst}}, \bibinfo {author} {\bibfnamefont {T.~B.}\ \bibnamefont
			{Schr{\o}der}},\ and\ \bibinfo {author} {\bibfnamefont {J.~C.}\ \bibnamefont
			{Dyre}},\ }\bibfield  {title} {\bibinfo {title} {Invariants in the {Yukawa}
			system's thermodynamic phase diagram},\ }\href@noop {} {\bibfield  {journal}
		{\bibinfo  {journal} {Phys. Plasmas}\ }\textbf {\bibinfo {volume} {22}},\
		\bibinfo {pages} {073705} (\bibinfo {year} {2015})}\BibitemShut {NoStop}%
	\bibitem [{\citenamefont {Rosenfeld}(1999)}]{ros99}%
	\BibitemOpen
	\bibfield  {author} {\bibinfo {author} {\bibfnamefont {Y.}~\bibnamefont
			{Rosenfeld}},\ }\bibfield  {title} {\bibinfo {title} {A quasi-universal
			scaling law for atomic transport in simple fluids},\ }\href@noop {}
	{\bibfield  {journal} {\bibinfo  {journal} {J. Phys.: Condens. Matter}\
		}\textbf {\bibinfo {volume} {11}},\ \bibinfo {pages} {5415} (\bibinfo {year}
		{1999})}\BibitemShut {NoStop}%
	\bibitem [{\citenamefont {Dyre}(2016)}]{dyr16}%
	\BibitemOpen
	\bibfield  {author} {\bibinfo {author} {\bibfnamefont {J.~C.}\ \bibnamefont
			{Dyre}},\ }\bibfield  {title} {\bibinfo {title} {Simple liquids'
			quasiuniversality and the hard-sphere paradigm},\ }\href
	{https://doi.org/10.1088/0953-8984/28/32/323001} {\bibfield  {journal}
		{\bibinfo  {journal} {J. Phys. Condens. Matter}\ }\textbf {\bibinfo {volume}
			{28}},\ \bibinfo {pages} {323001} (\bibinfo {year} {2016})}\BibitemShut
	{NoStop}%
	\bibitem [{\citenamefont {Thol}\ \emph {et~al.}(2015)\citenamefont {Thol},
		\citenamefont {Rutkai}, \citenamefont {Span}, \citenamefont {Vrabec},\ and\
		\citenamefont {Lustig}}]{thol2015equation}%
	\BibitemOpen
	\bibfield  {author} {\bibinfo {author} {\bibfnamefont {M.}~\bibnamefont
			{Thol}}, \bibinfo {author} {\bibfnamefont {G.}~\bibnamefont {Rutkai}},
		\bibinfo {author} {\bibfnamefont {R.}~\bibnamefont {Span}}, \bibinfo {author}
		{\bibfnamefont {J.}~\bibnamefont {Vrabec}},\ and\ \bibinfo {author}
		{\bibfnamefont {R.}~\bibnamefont {Lustig}},\ }\bibfield  {title} {\bibinfo
		{title} {Equation of state for the {Lennard-Jones} truncated and shifted model
			fluid},\ }\href@noop {} {\bibfield  {journal} {\bibinfo  {journal}
			{International Journal of Thermophysics}\ }\textbf {\bibinfo {volume} {36}},\
		\bibinfo {pages} {25} (\bibinfo {year} {2015})}\BibitemShut {NoStop}%
	\bibitem [{\citenamefont {Thol}\ \emph {et~al.}(2016)\citenamefont {Thol},
		\citenamefont {Rutkai}, \citenamefont {K{\"o}ster}, \citenamefont {Lustig},
		\citenamefont {Span},\ and\ \citenamefont {Vrabec}}]{thol2016equation}%
	\BibitemOpen
	\bibfield  {author} {\bibinfo {author} {\bibfnamefont {M.}~\bibnamefont
			{Thol}}, \bibinfo {author} {\bibfnamefont {G.}~\bibnamefont {Rutkai}},
		\bibinfo {author} {\bibfnamefont {A.}~\bibnamefont {K{\"o}ster}}, \bibinfo
		{author} {\bibfnamefont {R.}~\bibnamefont {Lustig}}, \bibinfo {author}
		{\bibfnamefont {R.}~\bibnamefont {Span}},\ and\ \bibinfo {author}
		{\bibfnamefont {J.}~\bibnamefont {Vrabec}},\ }\bibfield  {title} {\bibinfo
		{title} {Equation of state for the {Lennard-Jones} fluid},\ }\href@noop {}
	{\bibfield  {journal} {\bibinfo  {journal} {Journal of Physical and Chemical
				Reference Data}\ }\textbf {\bibinfo {volume} {45}},\ \bibinfo {pages}
		{023101} (\bibinfo {year} {2016})}\BibitemShut {NoStop}%
\end{thebibliography}
\end{document}